\documentclass[12pt,showpacs,showkeys,amsmath,amssymb]{revtex4}
\usepackage{amsmath,amsfonts,amsthm,amscd,amssymb,latexsym}
\usepackage{bm}
\usepackage{dcolumn}
\usepackage{graphicx}
\usepackage{epstopdf}
\usepackage{color}
\usepackage{epsf}
\usepackage{epsfig}
\usepackage{graphicx, epic, eepic, color}

\def\@{\partial_}

\def\negenspace{\kern-1.1em}

\def\sqr#1#2{{\vcenter{\hrule height.#2pt\hbox{\vrule width.#2pt
height#1pt \kern#1pt \vrule width.#2pt}\hrule height.#2pt}}}

\date{\today}

\begin{document}
\title{Mach's Principle and the Origin of Inertia}

\author{B. Mashhoon}
\email{mashhoonb@missouri.edu}
\affiliation{Department of Physics and Astronomy,
University of Missouri, Columbia, Missouri 65211, USA}

\begin{abstract} 

The current status of Mach's principle is discussed within the context of general relativity. The inertial
properties of a particle are determined by its mass and spin, since these characterize the irreducible unitary representations of the inhomogeneous Lorentz group. The origin of the inertia of mass and intrinsic spin are discussed and the inertia of intrinsic spin is studied via the coupling of intrinsic spin with rotation. The implications of spin-rotation coupling and the possibility of history dependence and nonlocality in relativistic physics are briefly mentioned. 

\end{abstract}



\maketitle

\section{Introduction}

Is motion absolute or relative? If the Newtonian absolute space and time are not real and therefore not responsible for the origin of inertia, then inertia must be due to immediate connections between masses. Thus one might expect that the inertial mass of a test particle increases in the vicinity of a large mass. Following this Machian line of thought, Einstein suggested that perhaps the physics of general relativity could be so interpreted as to allow for this possibility~\cite{Ei}. However, Brans showed that if one adopts the modern geometric interpretation of general relativity, then the inertial mass of a free test particle cannot change in a gravitational field. Indeed, this issue has since been completely settled as a result of Brans's thorough analysis~\cite{B1, B2}. Moreover, Mach's principle played an important role in the scalar-tensor generalization of Einstein's tensor theory by Brans and Dicke~\cite{BD}. 

Carl Brans has made basic contributions  to gravitation theory and general relativity.  \emph{It is a great pleasure for me to dedicate this paper to Carl on the occasion of his eightieth birthday.}  

The connection between Mach's ideas~\cite{MA} and Einstein's theory of gravitation~\cite{Ei} has been the subject of many interesting investigations~\cite{BaP}; furthermore, there is a diversity of opinion on this matter~\cite{BaP, MW, Ba, MaW}. A more complete account of the views expressed in this brief  treatment is contained in Ref.~\cite{LiM} and the references cited therein. 

The special theory of relativity involves, among other things,  the measurements of observers in Minkowski spacetime. The special class of \emph{inertial} observers has played a pivotal role in the development of physics, since the fundamental laws of physics are expressed in terms of the measurements of these \emph{hypothetical} observers. Indeed, inertial physics was originally established by Newton~\cite{Coh}. 
The measurements of inertial observers, each forever at rest in a given inertial frame in Minkowski spacetime, are related to each other by Lorentz invariance. Actual observers are all more or less accelerated. The measurements of accelerated observers must be interpreted in terms of the fundamental (but hypothetical) inertial observers. 

The acceleration of an observer in Minkowski spacetime is independent of any system of coordinates and is in this sense \emph{absolute}. Let $u^\mu=dx^\mu/d\tau$ be the 4-velocity of an observer following a timelike world line in Minkowski spacetime. Here, $x^\mu=(ct, \mathbf{x})$ denotes an event in spacetime, Greek indices run from 0 to 3, while Latin indices run from 1 to 3; moreover, the signature of the metric is +2 and $\tau$ is the proper time along the path of the observer. The observer's 4-acceleration $a^\mu= du^\mu/d\tau$ is orthogonal to $u^\mu$, namely, $a_\mu \, u^\mu =0$, since $u_\mu\, u^\mu = -1$. Thus $a^\mu$ is a spacelike 4-vector such that $a_\mu\, a^\mu = \mathcal{A}^2$, where the scalar $\mathcal{A}(\tau) \ge 0$ is the coordinate-independent magnitude of the observer's acceleration. An accelerated electric charge radiates electromagnetic waves. The existence of radiation is independent of inertial frames of reference and wave motion is in this sense absolute as well. 

Relativity theory, namely, Lorentz invariance, is extended to accelerated observers in Minkowski spacetime in a pointwise manner. That is, at each instant along its world line, the accelerated observer is assumed to be momentarily equivalent to a hypothetical inertial observer following the straight tangent world line at that instant. The further extension of relativity theory to observers in a gravitational field is accomplished via Einstein's \emph{local} principle of equivalence. Therefore,  in general relativity  spacetime is curved and the gravitational field is identified with spacetime curvature. Test particles and null rays follow timelike and null geodesics of the spacetime manifold, respectively. Moreover, spacetime is locally flat and Minkowskian. The global inertial frames of special relativity are thus replaced by local inertial frames; moreover, gravitation is rendered \emph{relative} in this way due to its universality, a circumstance that does not extend to other fundamental interactions. The general equation of motion of a classical point particle in general relativity  is given by
\begin{equation}\label{I}
\frac{d^2 x^\mu}{d\tau^2} + \Gamma^\mu_{\alpha \beta}\, \frac{d x^\alpha}{d\tau}\, \frac{d x^\beta}{d\tau} = a^\mu\,,
\end{equation}
where $a_\mu$ is the absolute acceleration of the particle due to nongravitational forces and $ \Gamma^\mu_{\alpha \beta}$ are the Christoffel symbols. For instance, for a particle of mass $m$ and charge $q$ in an electromagnetic field $F_{\mu \nu}$, $a_\mu = (q/m) F_{\mu \nu} \, u^\nu$ by the Lorentz force law. At each event in spacetime, coordinates can be chosen such that the Christoffel symbols all vanish and Eq.~\eqref{I} reduces to the corresponding relation in Minkowski spacetime. 

Inertial forces appear in a system (``laboratory") that accelerates with respect to the ensemble of \emph{local} inertial frames. These inertial forces are \emph{not} due to the gravitational influence of distant masses, which would instead generate gravitational tidal effects in the laboratory~\cite{MHT, BM, BaM1, BaM2, BaM3}.

\section{Mach's Principle}

Newton's \emph{absolute} space and time refer to the ensemble of \emph{inertial} frames, namely, Cartesian systems of reference that are homogeneous and isotropic in space and time and in which Newton's fundamental laws of motion are valid. Indeed, Newton's first law of motion, the principle  of inertia, essentially contains the definition of an inertial frame. Newton argued that the existence of \emph{inertial forces} provided observational proof of the reality of absolute space and time. Thus in classical mechanics, the motion of a Newtonian point particle is absolute, yet subject to the principle of relativity. However, the absolute motion of a particle, namely, its motion with respect to absolute space and time is not \emph{directly} observable. 

Mach considered \emph{all} motion to be relative and therefore argued against the Newtonian conceptions of absolute space and time~\cite{MA}. In his critique of Newtonian foundations of physics, Mach analyzed, among other things, the \emph{operational} definitions of time and space via masses and concluded that in Newtonian mechanics, masses are not organically connected to space and time~\cite{MA}. In fact, in chapter II of  Ref.~\cite{MA}, on pages 295-296 of section VI, we find:
\begin{quote}``\ldots Although I expect that astronomical observation will only as yet necessitate very small corrections, I consider it possible that the law of inertia in its simple Newtonian form has only, for us human beings, a meaning which depends on space and time. Allow me to make a more general remark.  We measure time by the angle of rotation of the
  earth, but could measure it just as well by the angle of rotation of
  any other planet. But, on that account, we would not believe that
  the {\em temporal} course of all physical phenomena would have to be
  disturbed if the earth or the distant planet referred to should
  suddenly experience an abrupt variation of angular velocity. We
  consider the dependence as not immediate, and consequently the
  temporal orientation as {\em external}. Nobody would believe that
  the chance disturbance --- say by an impact --- of one body in a system
  of uninfluenced bodies which are left to themselves and move
  uniformly in a straight line, where all the bodies combine to fix
  the system of co\"ordinates, will immediately cause a disturbance of
  the others as a consequence. The orientation is external here
  also. Although we must be very thankful for this, especially when it
  is purified from meaninglessness, still the natural investigator must
  feel the need of further insight --- of knowledge of the {\em immediate}
  connections, say, of the masses of the universe\ldots .''
\end{quote}

Thus Newton's absolute space and time are fundamentally different from their operational definitions by means of masses. Moreover, masses do not appear to ``recognize" absolute space and time, since they have been ``placed" in this arena without being immediately connected to it. In fact, the internal state of a Newtonian point particle, characterized by its mass $m$, has no direct connection with its external state in absolute space and time, characterized by its position and velocity $(\mathbf{x}, \mathbf{v})$ at a given  time $t$.  Thus only the \emph{relative} motion of classical particles is directly observable. This epistemological shortcoming of Newtonian mechanics  means that the external  state of the particle $m$, namely,  $(\mathbf{x}, \mathbf{v})$,  can in principle be occupied by other particles \emph{comoving} with it. It is indeed a prerequisite of the notion of \emph{relativity} of motion of masses that an observer be capable of changing its perspective by comoving with each particle in turn. As particles can be \emph{directly} connected with each other via interactions, we may say that particles have a \emph{propensity} for relative motion. 

With the advent of Maxwell's electrodynamics,  Galilean relativity was gradually replaced by Lorentz invariance, in which the speed of light is the same in all inertial frames of reference. Indeed, it is impossible for an inertial observer to be comoving with light. Motion is either relative or absolute in classical physics. The motion of light in Minkowski's absolute spacetime is independent of inertial observers and is, in this sense, \emph{nonrelative} or absolute.  Thus in Lorentz invariance, the motion of inertial observers is relative, while the motion of electromagnetic radiation is absolute. 

In classical physics, movement takes place via either particles or electromagnetic waves. As emphasized by Mach~\cite{MA}, the internal and external states of particles are not directly related, which leads to the notion of \emph{relativity} of particle motion. However, the opposite is the case for electromagnetic waves. That is, the internal state of a wave, namely, its period, wavelength, intensity and polarization are all directly related  to its external state characterized by its wave function, which is a solution of Maxwell's field equations. In this way, electromagnetic waves can ``recognize" absolute spacetime and this leads to their propensity for absolute motion. Therefore, an inertial frame of reference can be characterized by  standing electromagnetic radiation~\cite{HAL}. Indeed, ring lasers are now regularly employed in inertial guidance systems. Similarly, the wave nature of matter in quantum theory can be used to establish an inertial frame of reference. For instance, the rotation of the earth can be detected via superfluid helium~\cite{SBP}. Moreover, atom interferometers can function as sensitive inertial sensors, since they can measure acceleration and rotation to rather high precision~\cite{RKW, GBK, GLK, DK, DHSJK}.

\section{Duality of Absolute and Relative Motion}

Classical physics is the correspondence limit of quantum physics. It is therefore interesting to extend the quantum duality of classical particles and waves  to their motions as well. That is, the motion of a quantum particle has complementary classical aspects in relative and absolute movements~\cite{BM}. 

The epistemological shortcoming of Newtonian mechanics that was pointed out by Mach~\cite{MA}  essentially disappears in the quantum theory. That is, wave-particle duality makes it possible for a (quantum) particle to ``recognize" absolute spacetime; moreover, it is impossible for a classical observer to be comoving with the particle in conformity with Heisenberg's uncertainty principle. Consider, for instance, the nonrelativistic motion of a particle of mass $m$ in a potential $V$ according to the Heisenberg picture. The Hamiltonian is
\begin{equation}\label{1}
\hat H = \frac{1}{2\,m}\, \hat{p}^2 + V(\hat{\mathbf{x}})\,.
\end{equation}
In this ``particle" representation, the momentum operator of the particle is given by 
\begin{equation}\label{2}
\hat {\mathbf{p}} = m\, \frac{d\hat{\mathbf{x}}}{dt}\,, 
\end{equation}
so that the fundamental quantum condition, $[\hat{x}^j, \hat{p}^k]={\rm i}\,\hbar\,\delta_{jk}$, can be written as
\begin{equation}\label{3}
\left[\hat{x}^j,  \frac{d\hat{x}^k}{dt}\right] = {\rm i}\,\frac{\hbar}{m}\,\delta_{jk}\,.
\end{equation}
In sharp contrast to classical mechanics, the inertial mass of the particle is related, albeit in a statistical sense, to the observables corresponding to its position and velocity. This connection, through Planck's constant $\hbar$, disappears when $m \to \infty$. A macroscopically massive system behaves classically, since the perturbation experienced by the system due to any disturbance accompanying an act of observation would be expected to be negligibly small. Similarly, let $\hat{\mathbf{L}}$, $m\, \hat{L}^i = \epsilon_{ijk}\,\hat{x}^j \, \hat{p}^k$, be the specific orbital angular momentum operator of the particle; then, 
\begin{equation}\label{4}
[\hat{L}^j,  \hat{L}^k] ={\rm i}\,\frac{\hbar}{m}\,\epsilon_{jkn}\,\hat{L}^n\,.
\end{equation}
The mechanical laws of classical physics pertaining to translational and rotational inertia hold in a  certain average sense  in quantum theory as well. Moreover, in terms of the Schr\"odinger picture, the state of the particle is characterized by a wave function $\Psi( t, \mathbf{x})$ that satisfies the Schr\"odinger equation. This equation is explicitly dependent upon the particle's inertial mass $m$, thereby connecting the internal and external states of the particle in the ``wave" representation. 

In relativistic quantum theory, rotational inertia involves the intrinsic angular momentum of the particle as well, thus leading to the inertia of intrinsic spin.

\section{Inertia of Intrinsic Spin} 

Mass and spin describe the irreducible representations of the Poincar\'e group~\cite{Wi}.  The state of a particle in spacetime is thus characterized by its mass and spin, which determine the inertial properties of the particle. In quantum theory, therefore, the inertial characteristics of a particle are determined by its inertial mass~\cite{WSC, MO, HRW} as well as intrinsic spin~\cite{BMa, BMas, MNHS, MK}. 

To examine the inertia of intrinsic spin, we recall that the \emph{total} angular momentum operator is the generator of rotations; therefore, we expect that intrinsic spin would couple to the rotation of a frame of reference in much the same way as orbital angular momentum. This means that in a macroscopic body rotating in the positive sense with uniform angular velocity $\boldsymbol{\Omega}$, the spins of the constituent particles do not naturally participate in the rotation and all instead tend to stay essentially fixed with respect to the local inertial frame. Thus relative to the rotating body, we have in the nonrelativistic approximation for each spin vector $\hat{\boldsymbol{\sigma}}$, 
\begin{equation}\label{5}
\frac{d\hat{\sigma}^i}{dt} + \epsilon_{ijk}\, \Omega^j\, \hat{\sigma}^k = 0\,,
\end{equation}
since the spin vector appears to precess with angular velocity $-\boldsymbol{\Omega}$ with respect to observers at rest with the rotating body. The Hamiltonian corresponding to this motion is
\begin{equation}\label{6}
\hat{{\cal H}}= -\hat{\sigma}\cdot \boldsymbol{\Omega}\,,
\end{equation}
because the Heisenberg equation of motion 
\begin{equation}\label{7}
{\rm i}\,\hbar\, \frac{d\hat{\sigma}^k}{dt}=[\hat{\sigma}^k,  \hat{{\cal H}}] 
\end{equation}
coincides with Eq.~\eqref{5}. For a discussion of the corresponding relativistic treatment and further developments of this subject, see Refs.~\cite{HN, LR, SP, Pa, Ran, JN}. Moreover, a review of this subject and a more complete list  of references is contained in Ref.~\cite{MASH}. 

In general, the energy of an incident particle as measured by the rotating observer is given by
\begin{equation}\label{8}
 E'=\gamma \,(E-\hbar\,M\Omega)\,,
\end{equation}
where $E$ is the energy of the incident particle in the inertial frame and  $M$ is the total (orbital plus spin) ``magnetic" quantum number along the axis of
rotation. In fact,  $M=0,\pm 1,\pm 2,\dots $, for a scalar or a vector particle, while $M\mp
\frac{1}{2} =0,\pm 1, \pm 2,\dots$, for a Dirac particle. 
In the JWKB approximation, $E'=\gamma\,(E-\boldsymbol{\Omega}\cdot \mathbf{J})$, 
where $\mathbf{J}=\mathbf{r} \times \mathbf{P} + \boldsymbol{\sigma}$
 is the total angular momentum of the particle and $\mathbf{P}$ is its momentum; hence, 
 $E'=\gamma\,(E-\mathbf{v}\cdot\mathbf{P})-\gamma\, \boldsymbol{\sigma}\cdot \boldsymbol{\Omega}$, where 
 $\mathbf{v}=\boldsymbol{\Omega}\times \mathbf{r}$ is the velocity of the uniformly rotating observer with respect to the background inertial frame and $\gamma$ is the Lorentz factor of the observer. The energy corresponding to spin-rotation coupling is naturally augmented by time dilation. 
 
 It is important to remark here that the spin-rotation coupling is completely independent of the inertial mass of the particle. Moreover, the associated spin-gravity  coupling is an interaction of the intrinsic spin with the gravitomagnetic field of the rotating source that is also independent of the mass of the test particle. For instance, free neutral Dirac particles with their spins up and down (i.e., parallel and antiparallel to the vertical direction, respectively) in general fall differently in the gravitational field of the rotating earth~\cite{BMas}.

The spin-rotation coupling has recently been measured for neutrons via neutron polarimetry~\cite{DSH}.
Moreover, this general coupling has now been incorporated into the condensed-matter physics of spin mechanics and spin currents~\cite{MISM, MISM2, MIHSM, MIM, IMM, CB, JQS, LM, HYM}. It is also expected to play a role in the emerging field of spintronics~\cite{PAP}.

\section{Helicity-Rotation Coupling}

 To illustrate the general nature of spin-rotation coupling, we now turn to the case of photons, see Refs.~\cite{PJA, Mash, BMash, BMash2, BGKH, KYB, BA} and the references cited therein.  Consider the measurement of the frequency of a plane monochromatic electromagnetic wave of frequency $\omega$ propagating along the axis of rotation of the observer. The result of the Fourier analysis of the measured field is
\begin{equation}\label{9}
 \omega '=\gamma \,(\omega \mp\Omega)\,,
\end{equation}
where the upper (lower) sign refers
to positive (negative) helicity radiation.  With $E=\hbar\, \omega$, our classical result~\eqref{9} is consistent with the general formula~\eqref{8} for spin 1 photons.  The
helicity-dependent  contribution to the transverse Doppler effect in Eq.~\eqref{9} has been verified via the GPS~\cite{NAS}, where it is responsible  for the phenomenon of  \emph{phase
wrap-up}~\cite{NAS}. 

It is simple to interpret the coupling of helicity with rotation in Eq.~\eqref{9}, aside from the presence of the Lorentz factor that
is due to time dilation.  In a positive (negative) helicity wave, the electromagnetic field rotates with
frequency $\omega$ ($-\omega)$ about the direction of propagation of the wave.  The
rotating observer therefore perceives positive (negative) helicity radiation with the
electromagnetic field rotating with frequency $\omega -\Omega$ ($-\omega-\Omega$) about
the direction of wave propagation.  

For the case of oblique incidence, the analog of Eq.~\eqref{9} is 
\begin{equation}\label{10}
 \omega '=\gamma \,(\omega - M\,\Omega)\,,
\end{equation}
where $M=0,\pm 1,\pm 2,\dots $ for the electromagnetic field. This exact classical result can be obtained by studying the electromagnetic field as measured by uniformly rotating observers.  It is interesting to  note that $\omega'=0$ for $\omega=M\,\Omega$, a situation that is discussed in the next section, while $\omega'$ can be negative for $\omega < M\,\Omega$. The latter circumstance does not pose any basic difficulty, since it is simply a consequence of the absolute character of rotational motion. 

\section{Can Light Stand Completely Still?}

The exact formula $ \omega '=\gamma\, (\omega - \Omega)$ for incident positive-helicity radiation has a remarkable consequence that is not easily accessible to experimental physics: The incident wave stands completely still with respect to observers that rotate uniformly with frequency $\Omega=\omega$ about the direction of propagation of the wave.  That is, helicity-rotation coupling has the consequence that a rotating observer can in principle be comoving with an electromagnetic wave; in fact, the wave appears to be oscillatory in space but stands completely still with respect to the rotating observer. The fundamental   difficulty under consideration here is quite general, as it occurs for oblique incidence as well. 

By a mere rotation, an observer can in principle stay completely at rest with respect to an electromagnetic wave. This circumstance is rather analogous to the difficulty with the pre-relativistic Doppler formula, where an inertial observer moving with speed $c$ along a light beam would see a wave that is oscillatory in space but is otherwise independent of time and hence completely at rest. This issue, as is well known, played a part in Einstein's path to relativity, as mentioned in his autobiographical notes, see page 53 of Ref.~\cite {Sch}.
The difficulty in that case was eventually removed by Lorentz invariance; however, in the present case, the problem has to do with the way Lorentz invariance is extended to accelerated observers in Minkowski spacetime. In the special theory of relativity, Lorentz invariance is extended to accelerated systems via the \emph{hypothesis of locality}, namely, the assumption that an accelerated observer is pointwise inertial~\cite{BM1, BM2}. This circumstance extends to general relativity through Einstein's \emph{local} principle of equivalence. The locality assumption originates from Newtonian mechanics, where the state of a particle is determined by $(\mathbf{x}, \mathbf{v})$ at time $t$. The accelerated observer shares this state with a comoving inertial observer; therefore, they are at that moment physically equivalent insofar as all physical phenomena could be reduced to pointlike coincidences of classical particles and rays of radiation. However, classical wave phenomena are in general intrinsically \emph{nonlocal}.

According to the locality assumption, the world line of an accelerated observer in Minkowski spacetime can be replaced at each instant by its tangent and then Lorentz transformations can be pointwise employed to determine what the accelerated observer measures. To go beyond the locality postulate of special relativity theory, the past history of the observer must be taken into account. Thus the locality postulate must be supplemented by a certain average over the past world line of the observer. In this way, the observer retains the memory of its past acceleration.  This averaging procedure involves a weight function that must be determined. In this connection, we introduce the fundamental assumption that \emph{a basic radiation field can never stand completely still with respect to an observer}. On this basis a nonlocal theory of accelerated observers can be developed~\cite{BM3}. The nonlocal approach is in better correspondence with quantum theory than the standard treatment based on the hypothesis of locality~\cite{Mashhoon}. 

History-dependent theories are nonlocal in the sense that the usual partial differential equations for the fields are replaced by integro-differential equations. Acceleration-induced nonlocality in Minkowski spacetime suggests that gravity could be nonlocal. This is due to the intimate connection between inertia and gravitation, implied by Mach (e.g., in his discussion of Newton's experiment with the rotating bucket of water on page 279 of Ref.~\cite{MA}) and developed in new directions by Einstein in his general theory of relativity~\cite{Ei}. That is, Einstein interpreted the principle of equivalence of inertial and gravitational masses to mean that an intimate connection exists between inertia and gravitation. One can follow Einstein's interpretation without postulating a \emph{local} equivalence between inertia and gravitation as in general relativity; for instance, one can instead extend general relativity to make it history dependent. Recently, nonlocal theories of special and general relativity have been developed~\cite{M1, HM1, HM2, M2, RM}. It turns out that nonlocal general relativity simulates dark matter. That is, according to this theory, what appears as dark matter in astrophysics~\cite{RF, RW, SR} is essentially a manifestation of the nonlocality of the gravitational interaction~\cite{RM}. 

\section{Discussion}

Classical relativistic mechanics and classical electrodynamics are mainly concerned with two types of motion, namely,  local particle motion and nonlocal wave motion, respectively. These are brought together in geometric optics, where the \emph{waves} are replaced by \emph{rays} that can be treated in a similar way as classical point particles. With respect to Minkowski's absolute spacetime, particle motion is absolute; however, this absolute motion is not directly observable. On the other hand, the motion of classical particles naturally tends to be \emph{relative}. Similarly,  the motion of electromagnetic waves naturally tends to be  \emph{absolute}, though the corresponding wave equation is Lorentz invariant. In the quantum domain, this line of thought leads to the complementarity of absolute and relative motion; moreover, the notion of inertia must be extended to include intrinsic spin as well.  The inertial coupling of intrinsic spin to rotation has recently been measured in neutron polarimetry~\cite{DSH}. The implications of the inertia of intrinsic spin are critically examined in the light of the hypothesis that an electromagnetic wave cannot stand completely still with respect to any accelerated observer.


\begin{thebibliography}{xxxx}

\bibitem{Ei}
 A. Einstein, \emph{The Meaning of Relativity} (Princeton University Press,
Princeton, NJ, 1955).

\bibitem{B1}
C. H. Brans, Phys. Rev. {\bf 125}, 388 (1962).

\bibitem{B2}
 C. H. Brans, Phys. Rev. Lett. {\bf 39}, 856 (1977).

\bibitem{BD} 
C. H. Brans and R. H. Dicke, Phys. Rev. {\bf 124}, 925 (1961).

\bibitem{MA}
 E. Mach, \emph{The Science of Mechanics} (Open Court, La Salle, 1960).
 

\bibitem{BaP}
J. Barbour and H. Pfister, editors, \emph{Mach's Principle: From Newton's Bucket
      to Quantum Gravity}, Einstein Studies {\bf 6} (Birkh\"auser, Boston, 1995).


\bibitem{MW}
B. Mashhoon and P. S. Wesson, ``Mach's Principle and Higher-Dimensional Dynamics", Ann. Phys. (Berlin) {\bf 524}, 63 (2012).
[arXiv: 1106.6036 [gr-qc]]


\bibitem{Ba}
J. Barbour, Ann. Phys. (Berlin) {\bf 524}, A39 (2012).
 [arXiv: 1108.3057 [gr-qc]]

\bibitem{MaW}
B. Mashhoon and P.S. Wesson, Ann. Phys. (Berlin) {\bf 524}, A44 (2012).
[arXiv: 1108.3059 [gr-qc]]

\bibitem{LiM}
H. Lichtenegger and B. Mashhoon, ``Mach's Principle", 
        in \emph{The Measurement of Gravitomagnetism: A Challenging Enterprise}, edited by L. Iorio 
        (NOVA Science, Hauppage, NY, 2007), Chapter 2.
        [arXiv: physics/0407078 [physics.hist-ph]]  


\bibitem{Coh}
I. B. Cohen, \textit{The Birth of a New Physics}  (Doubleday Anchor Books, Garden City, NY, 1960). 

\bibitem{MHT} 
 B. Mashhoon, F. W. Hehl and D. S. Theiss, Gen. Relativ. Gravit. {\bf 16}, 711 (1984).
 
\bibitem{BM} 
B. Mashhoon, ``Complementarity of Absolute and Relative Motion", Phys. Lett. A {\bf 126}, 393 (1988).

\bibitem{BaM1}
 B. Mashhoon, Found. Phys. Lett. {\bf 6}, 545 (1993).
 
 \bibitem{BaM2}
  B. Mashhoon, in {\em Directions in General Relativity: Papers in Honor of
 Dieter Brill}, edited by B. L. Hu and  T. A. Jacobson  (Cambridge University Press, Cambridge, UK, 1993),  pp. 182-194.

\bibitem{BaM3}
B. Mashhoon, Nuovo Cimento B {\bf 109}, 187 (1994).

\bibitem{HAL}
H. A. Lorentz, Nature {\bf 112}, 103 (1923). 

\bibitem{SBP} K. Schwab, N. Bruckner and R. E. Packard, Nature {\bf 386}, 585 (1997).
 
\bibitem{RKW}
F. Riehle, Th. Kisters, A. Witte, J. Helmcke and Ch. J. Bord\'e, Phys. Rev. Lett. {\bf 67}, 177 (1991).

\bibitem{GBK}
T. L. Gustavson, P. Bouyer and M. A. Kasevich, Phys. Rev. Lett. {\bf 78}, 2046 (1997). 

\bibitem{GLK}
T. L. Gustavson, A. Landragin and M. A. Kasevich, Classical Quantum Gravity {\bf 17}, 2385 (2000).

\bibitem{DK}
B. Dubetsky and M. A. Kasevich, Phys. Rev. A {\bf 74}, 023615 (2006).

\bibitem{DHSJK}
S. M. Dickerson, J. M. Hogan, A. Sugarbaker, D. M. S. Johnson and M. A. Kasevich, 
Phys. Rev. Lett. {\bf 111}, 083001 (2013). 

\bibitem{Wi} E. P. Wigner, Ann. Math. {\bf 40}, 149 (1939).

\bibitem{WSC} 
S. A. Werner, J.-L. Staudenmann and R. Collela, Phys. Rev. Lett. {\bf 42}, 1103 (1979).

\bibitem{MO} 
G. F. Moorhead and G. I. Opat, Classical Quantum Gravity {\bf 13}, 3129 (1996).

\bibitem{HRW}
H. Rauch and S. A. Werner, \emph{Neutron Interferometry}, 2nd edition (Oxford University Press, Oxford, UK, 2015).

\bibitem{BMa}
 B. Mashhoon, Phys. Rev. Lett. {\bf 61}, 2639 (1988).

\bibitem{BMas} 
B. Mashhoon, Phys. Lett. A {\bf 198}, 9 (1995).

\bibitem{MNHS} 
B. Mashhoon, R. Neutze, M. Hannam and G. E. Stedman, Phys. Lett. A {\bf 249}, 161 (1998).

\bibitem{MK}
B. Mashhoon and H. Kaiser, ``Inertia of Intrinsic Spin", Physica B {\bf 385-386}, 1381 (2006).
[arXiv: quant-ph/0508182]

\bibitem{HN}
 F. W. Hehl and W.-T. Ni, Phys. Rev. D {\bf 42}, 2045 (1990).

\bibitem{LR} 
L. Ryder, J. Phys. A {\bf 31}, 2465 (1998).

\bibitem{SP}
D. Singh and G. Papini, Nuovo Cimento B {\bf 115}, 223 (2000).

\bibitem{Pa} 
G. Papini, Phys. Rev. D {\bf 65}, 077901 (2002).


\bibitem{Ran}
A. Randono, Phys. Rev. D {\bf 81}, 024027 (2010). 

\bibitem{JN}
U. D. Jentschura and J. H. Noble, J. Phys. A: Math. Theor.  {\bf 47}, 045402 (2014).


\bibitem{MASH}
 B. Mashhoon, Lect. Notes Phys. \textbf{702}, 112 (2006).      [arXiv: hep-th/0507157]

\bibitem{DSH}
B. Demirel, S. Sponar and Y. Hasegawa, ``Measurement of the Spin-Rotation Coupling in Neutron Polarimetry", New J. Phys. {\bf 17}, 023065 (2015).


\bibitem{MISM}
M. Matsuo, J. Ieda, E. Saitoh and S. Maekawa, ``Effects of Mechanical Rotation on Spin Currents", Phys. Rev. Lett. {\bf 106},  076601 (2011). 

\bibitem{MISM2}
M. Matsuo, J. Ieda, E. Saitoh and S. Maekawa, ``Spin-Dependent Inertial Force and Spin Current in Accelerating Systems", Phys. Rev. B {\bf 84},  104410 (2011). 

\bibitem{MIHSM}
M. Matsuo, J. Ieda, K. Harii, E. Saitoh and S. Maekawa, ``Mechanical Generation of Spin Current by Spin-Rotation Coupling", Phys. Rev. B {\bf 87},  180402(R) (2013). 

\bibitem{MIM}
M. Matsuo, J. Ieda  and S. Maekawa, ``Renormalization of Spin-Rotation Coupling", Phys. Rev. B {\bf 87},  115301 (2013). 

\bibitem{IMM}
J. Ieda, M. Matsuo  and S. Maekawa, ``Theory of Mechanical Spin Current Generation via Spin-Rotation Coupling", Solid State Communications {\bf 198},  52  (2014). 

\bibitem{CB}
D. Chowdhury and B. Basu, ``Effect of Spin Rotation Coupling on Spin Transport", Ann. Phys. {\bf 339}, 358 (2013). 

\bibitem{JQS}
J.-Q. Shen and S.-L. He, ``Geometric Phases of Electrons due to Spin-Rotation Coupling in Rotating C60 Molecules", Phys. Rev. B {\bf 68}, 195421 (2003).

\bibitem{LM}
J. R. F. Lima and F. Moraes, ``The Combined Effect of Inertial and Electromagnetic Fields in a Fullerene Molecule", Eur. Phys. J. B {\bf 88}, 63 (2015).

\bibitem{HYM}
M. Hamada, T. Yokoyama and S. Murakami, ``Spin Current Generation and Magnetic Response in Carbon Nanotubes by the Twisting Phonon Mode", Phys. Rev. B {\bf 92}, 060409(R) (2015). 

\bibitem{PAP}
G. Papini, ``Spin Currents in Non-Inertial Frames", Phys. Lett. A {\bf 377}, 960 (2013). 


\bibitem{PJA} 
P. J. Allen, Am. J. Phys. {\bf 34}, 1185 (1966). 

\bibitem{Mash} 
B. Mashhoon, Phys. Lett. A {\bf 139}, 103 (1989). 
 
\bibitem{BMash} 
B. Mashhoon, ``Optics of Rotating Systems", Phys. Rev. A {\bf 79}, 062111 (2009). 
[arXiv: 0903.1315 [gr-qc]]
 
\bibitem{BMash2}
B. Mashhoon, ``Nonlocal Special Relativity: Amplitude Shift in Spin-Rotation Coupling",
in  \emph{Proc.  Mario Novello's 70th Anniversary Symposium}, edited by N. Pinto Neto and S. E. Perez Bergliaffa (Editora Livraria da Fisica, Sao Paulo, 2012), pp. 177-189. 
[arXiv:1204.6069 [gr-qc]]

\bibitem{BGKH}
K. Y. Bliokh, Y. Gorodetski, V. Kleiner and E. Hasman, Phys. Rev. Lett. {\bf 101}, 030404 (2008). 

\bibitem{KYB}
K. Y. Bliokh, J. Opt. A: Pure Appl. Opt. {\bf 11}, 094009 (2009). 
 
\bibitem{BA}
K. Y. Bliokh and A. Aiello, J. Opt. {\bf 15}, 014001 (2013).  
 
\bibitem{NAS} 
N. Ashby,  ``Relativity in the Global Positioning System", Living Rev. Relativity {\bf 6}, 1 (2003). 
\url{http://www.livingreviews.org/lrr-2003-1}
 
\bibitem{Sch} 
P. A. Schilpp, \emph{Albert Einstein: Philosopher-Scientist} (Library of Living Philosophers, Evanston, Illinois, 1949). 

\bibitem{BM1} 
B. Mashhoon, ``The Hypothesis of Locality in Relativistic Physics",  Phys. Lett. A \textbf{145}, 147 (1990).

\bibitem{BM2} 
B. Mashhoon, ``Limitations of Spacetime Measurements",  Phys. Lett. A \textbf{143}, 176 (1990).

\bibitem{BM3}
B. Mashhoon, ``Nonlocal Theory of Accelerated Observers", Phys. Rev. A \textbf{47}, 4498 (1993). 
 
\bibitem{Mashhoon} 
B. Mashhoon, Phys. Rev. A \textbf{72}, 052105 (2005).   [arXiv: hep-th/0503205]

\bibitem{M1}
B. Mashhoon, ``Nonlocal Special relativity", Ann. Phys. (Berlin) {\bf  17},  705 (2008).
[arXiv: 0805.2926 [gr-qc]]

\bibitem{HM1}
F.~W.~Hehl and B.~Mashhoon, ``Nonlocal Gravity Simulates Dark Matter", 
Phys. Lett. B {\bf 673}, 279 (2009).   [arXiv: 0812.1059 [gr-qc]]


\bibitem{HM2} 
F.~W.~Hehl and B.~Mashhoon, ``Formal Framework for a Nonlocal Generalization of Einstein's Theory of Gravitation", 
Phys. Rev. D {\bf 79}, 064028 (2009).   [arXiv: 0902.0560 [gr-qc]] 

\bibitem{M2}  
B.~Mashhoon, ``Nonlocal Gravity: The General Linear Approximation", Phys. Rev. D {\bf 90}, 124031 (2014).   [arXiv:1409.4472 [gr-qc]]      
          
 
\bibitem{RM} 
S.~Rahvar and B.~Mashhoon, ``Observational Tests of Nonlocal Gravity: Galaxy Rotation Curves and Clusters of Galaxies", Phys. Rev. D {\bf 89}, 104011 (2014).   [arXiv:1401.4819 [gr-qc]]

 \bibitem{RF} 
V. C. Rubin and W. K. Ford, Astrophys. J.  {\bf 159}, 379 (1970).  

\bibitem{RW}
 M. S. Roberts and R. N. Whitehurst, Astrophys. J.  {\bf 201}, 327 (1975).

\bibitem{SR} 
Y. Sofue and V. Rubin, ``Rotation Curves of Spiral Galaxies", 
Annu. Rev. Astron. Astrophys. {\bf 39}, 137 (2001).



   
\end{thebibliography}
\end{document}